\documentclass[prd,twocolumn, nofootinbib,superscriptaddress,preprintnumbers]{revtex4}

\usepackage{placeins}
\usepackage{soul}
\usepackage{chngcntr}
\usepackage{color}
\usepackage{epsfig}  
\usepackage{graphicx}
\usepackage{tabularx}
\usepackage{xspace}
\usepackage{float}
\usepackage{comment}
\usepackage[usenames, dvipsnames]{xcolor}

\usepackage{color}
\usepackage{hyperref}
\hypersetup{
     colorlinks   = true,
     citecolor    = MidnightBlue,
	 linkcolor=MidnightBlue
}
\usepackage{halloweenmath}
\usepackage{verbatim}
\usepackage{amsmath}
\usepackage{amssymb}
\usepackage{mathtools}
\usepackage{url}
\usepackage{bbold}
\usepackage{slashed}
\usepackage{array}
\usepackage{comment}

\usepackage{multirow}
\usepackage{threeparttable}
\usepackage{paralist}
\usepackage{xspace}
\usepackage{upgreek}
\usepackage{lipsum}
\usepackage{mathrsfs}

\newcommand\colvec[3][]{\begin{pmatrix}\ifx\relax#1\relax\else#1\\\fi#2\\#3\end{pmatrix}}

\definecolor{darkmagenta}{rgb}{0.55, 0.0, 0.55}
\newcommand{\beq}{\begin{equation}}
\newcommand{\beqn}{\begin{eqnarray}}
\newcommand{\eeq}{\end{equation}}
\newcommand{\eeqn}{\end{eqnarray}}

\newcommand{\order}[1]{\mathcal{O}{(#1)}}
\newcommand{\nl}{\nonumber \\}

\DeclareRobustCommand{\Eq}[1]{Eq.~\eqref{#1}}
\DeclareRobustCommand{\Eqs}[2]{Eqs.~\eqref{#1} and \eqref{#2}}

\DeclareRobustCommand{\Sec}[1]{Section~\ref{#1}}

\DeclareRobustCommand{\Fig}[1]{Fig.~\ref{#1}}

\DeclareMathAlphabet\mathbfcal{OMS}{cmsy}{b}{n}
 
\renewcommand{\vec}[1]{\mathbf{#1}}

\newcommand{\eV}{ \textrm{eV} }

\newcommand{\km}{ \textrm{km} }

\newcommand{\w}{\omega}

\newcommand{\Ap}{A^\prime}
\newcommand{\mAp}{m_{A^\prime}}
\newcommand{\meff}{m_\text{eff}}
\newcommand{\eps}{\epsilon}

\newcommand{\lrarr}{\leftrightarrow}

\newcommand{\prob}{P_{A_a \leftrightarrow A_s}}

\newcommand{\mutitle}{resonance enhancement parameter }

\begin{document}
\preprint{FERMILAB-PUB-23-421-T}

\title{Photon-Dark Photon Conversion with Multiple Level Crossings}
\author{Nirmalya Brahma}
\affiliation{Department of Physics \& Trottier Space Institute, McGill University, Montr\'eal, QC H3A 2T8, Canada}
\author{Asher Berlin}
\affiliation{Theory Division, Fermi National Accelerator Laboratory, Batavia, IL 60510, USA }
\affiliation{Superconducting Quantum Materials and Systems Center (SQMS),
Fermi National Accelerator Laboratory, Batavia, IL 60510, USA}
\affiliation{Perimeter Institute for Theoretical Physics, Waterloo, Ontario N2L 2Y5, Canada}
\author{Katelin Schutz}
\affiliation{Department of Physics \& Trottier Space Institute, McGill University, Montr\'eal, QC H3A 2T8, Canada}

\begin{abstract}\noindent

\noindent
Dark photons can oscillate into Standard Model (SM) photons via kinetic mixing. The conversion probability depends sensitively on properties of the ambient background, such as the density and electromagnetic field strength, which cause the SM photon to acquire an in-medium effective mass. Resonances can enhance the conversion probability when there is a level-crossing between the dark photon and background-dependent SM photon states. In this work, we show that the widely used Landau-Zener (LZ) approximation breaks down when there are multiple level-crossings due to a non-monotonic SM photon potential. Phase interference effects, especially when the dark photon mass is close to an extremum of the SM photon effective mass, can cause deviations from the LZ approximation at the level of a few orders of magnitude in the conversion probability. We present an analytic approximation that is valid in this regime and that can accurately predict the conversion probabilities in a wide range of astrophysical environments.
\end{abstract}
\maketitle
\section{Introduction}
Dark photons (DPs) are the gauge bosons of a hidden $U(1)'$ symmetry that may be spontaneously broken (analogous to the Higgs mechanism) or broken explicitly (with the Stuckelberg action for a spin-1 field), yielding a non-zero dark photon mass. In general, DPs will mix with Standard Model (SM) photons at some level via the dimension-4 kinetic mixing operator. The additional terms in the Lagrangian capturing the effects of the DP field $A'$ mixing with the SM photon field $A$ are
\beq
\label{Lagrangian}
    \mathcal{L} \supset  -\frac{1}{4}\, F'_{\mu\nu} \, F'^{\mu\nu} +\frac{\eps}{2} \, F'_{\mu\nu} \, F^{\mu\nu}  +
    \frac{1}{2} \, \mAp^{2} \, A'_{\mu} \, {A}'^{\mu} ~,
\eeq
where $F'$ and $F$ are the dark and SM field strength tensors, $\eps \ll 1$ is the dimensionless coupling that determines the strength of kinetic mixing with the SM photon, and $\mAp$ is the DP mass. Kinetic mixing has attracted considerable interest as a portal to dark sectors that would be accessible at a wide range of energies since it is a marginal operator. In particular, $\eps$ could be generated by any number of mechanisms~\cite{Holdom:1985ag,Dienes:1996zr,Abel:2003ue,Batell:2005wa,Aldazabal:2000sa,Abel:2004rp,Abel:2008ai,Acharya:2016fge,Acharya:2017kfi,Gherghetta:2019coi,Arkani-Hamed:2008kxc}, for instance by loop diagrams with heavy matter fields charged under both $U(1)'$ and the SM $U(1)$, or alternatively from certain realizations of string theory. From a bottom-up effective field theory point of view, $\eps$ can be thought of as a free parameter accompanying a marginal operator.

Astrophysical and cosmological observations provide some of the tightest constraints on the existence of DPs over a wide range of masses~\cite{Fischbach:1994ir,Vinyoles_2015,Redondo:2013lna,An:2013yfc,McDermott:2019lch,Mirizzi:2009iz,Caputo:2020bdy,Caputo:2020rnx,Li:2023vpv,Garcia:2020qrp}. These constraints often involve oscillations between DPs and SM photons leading to spectral imprints or new energy loss mechanisms. Furthermore, this conversion is resonantly enhanced if there is an energy level-crossing due to in-medium contributions to the SM photon effective mass. Large corrections to the SM photon mass arise from ambient free charges or strong background electromagnetic fields~\cite{kapusta2006finite,1936ZPhy...98..714H}, and corrections to the DP effective mass can also arise due to a background of dark sector particles directly charged under the DP~\cite{Berlin:2022hmt,Berlin:2023gvx}. 

In-medium effective masses can vary considerably over spatial or temporal domains in astrophysical and cosmological environments. In the case of non-adiabatic transitions due to a level-crossing, the conversion probability is usually well-approximated by the Landau-Zener (LZ) formula~\cite{Zener:1932ws,Landau:1932vnv}, which is often employed in such calculations as the leading-order term in the stationary phase approximation. However, this formalism breaks down when applied to the special case of resonant conversions occurring near local minima or maxima of the SM photon's in-medium mass (which can depend on the photon frequency). This breakdown therefore occurs generically in environments where the density of SM particles is non-monotonic in space or time as traversed by SM photons. The conversion probability predicted by the LZ formula can deviate from the full solution to the Schr\"odinger equation (to leading order in $\eps$) by several orders of magnitude. Therefore, there may be regions of DP parameter space at specific masses where existing constraints are subject to large corrections. In this work, we provide a simple analytic expression, \Eq{prob_crit}, that is more generally applicable to this region of parameter space and discuss the potential impact on DP oscillation probabilities. We note that although we concretely work with DP oscillations in this paper, the formalism developed here can potentially be applied to axion-photon conversions as well as neutrino oscillations.

The rest of this paper is organized as follows. In \Sec{sec:oscill}, we review various formalisms for computing transition probabilities for two-state systems. In \Sec{sec:toymodel}, we quantify the breakdown of the LZ approximation near the critical mass with an illustrative toy example. In \Sec{sec:astro}, we examine the consequences of the breakdown of the LZ approximation in a few case examples arising in astrophysical and cosmological settings, such as neutron star magnetospheres and the solar chromosphere, as well as during the epoch of reionization. Concluding remarks follow in \Sec{sec:conclusion}. 

\section{Photon-Dark photon oscillation formalism}
\label{sec:oscill}

\subsection{Photon-dark photon oscillations in an arbitrary in-medium potential}
The kinetic terms in \Eq{Lagrangian} can be made canonical by moving to the active-sterile basis described by the $A_{a}$ and $A_{s}$ fields,
\beq
\boldsymbol{A}^\mu \equiv \begin{pmatrix}A_{a}^\mu\\ A_{s}^\mu\end{pmatrix}=
\begin{pmatrix}1 & 0\\ -\eps &1\end{pmatrix}\begin{pmatrix}A^\mu \\ A^{\prime \mu}\end{pmatrix}
+\order{\eps^2}
~,
\eeq
to leading order in $\eps$. Using this in \Eq{Lagrangian}, we then have
\begin{align}
 \label{dp_lagrangian}
     \mathcal{L} &=-\frac{1}{4} \, F_{\mu\nu}^{a} \, F^{\mu\nu}_{a} - \frac{1}{4} \, F_{\mu\nu}^{s} \, F^{\mu\nu}_{s} + e \, J_{\mu}\, A_{a}^{\mu}
     \nl
     & + \frac{1}{2} \boldsymbol{A}_\mu^T \begin{pmatrix} 0 & \eps \, \mAp^2\\ \eps \, \mAp^2& \mAp^2\end{pmatrix}
    \boldsymbol{A}^\mu 
     +\order{\eps^2}
     ~,
\end{align}
where $F_{a}^{\mu\nu}$ and $F_{s}^{\mu\nu}$ are the active and sterile field strengths and we have also included the coupling of $A_a$ to the SM current density $J$. This basis is often referred to as the interaction basis, since SM currents selectively couple to  the active state $A_a$. However, note that the mass matrix is non-diagonal in this case. As a result, although the sterile state is not sourced directly from SM currents, it arises indirectly from $A_a \leftrightarrow A_s$ oscillations, which are the focus of this work.

In a medium, the free charges and electromagnetic fields in the background can source a potential that alters the dispersion relation of the active photons. We parameterize this effect by including a spatially-dependent in-medium  mass $\meff (\vec{x})$ for the visible field $A_a$ in \Eq{dp_lagrangian}. The in-medium mass-squared matrix is then given by
\beq
\label{in-medium_mass}
    \mathcal{M}^{2} (\vec{x}) \simeq \begin{pmatrix} \meff^2 (\vec{x}) & \eps \, \mAp^2\\ \eps \, \mAp^2& \mAp^2\end{pmatrix}
    ~.
\eeq
In this Section, we remain agnostic about the specific form of $\meff (\vec{x})$, leaving an exploration of specific examples to \Sec{sec:astro}.

We can track the propagation of the $A_a - A_s$ system by solving the corresponding equation of motion from \Eqs{dp_lagrangian}{in-medium_mass}. Following Ref.~\cite{Raffelt:1987im}, we switch to Fourier space, where $\w$ and $k$ are the frequency and wavenumber of the field $\boldsymbol{A}^\mu$, respectively, and assume that $\meff (\vec{x})$ varies on scales much larger than $k^{-1}$. In this case, we can approximate this in-medium contribution as a constant {on the scale of the de Broglie wavelength}, such that the equation of motion for transverse modes takes the form of a standard wave equation, 
\beq
\label{2ndpropagation}
   \left[ \w^{2} -k^2 -\mathcal{M}^{2} (\vec{x})   \right] \boldsymbol{A}^\mu (\w, k) = 0~.
\eeq
We note that an analogous dispersion relation of the form $( \w^{2} -\mathcal{M}_L^{2} ) \, \boldsymbol{A}_L = 0$ holds for longitudinal modes, but a dedicated study of their evolution is beyond the scope of this work. 

To proceed, we expand \Eq{2ndpropagation} in the relativistic limit $\w \simeq k \gg \mAp, \meff$ to obtain a linearized Schr\"{o}dinger-like equation, as in Ref.~\cite{Raffelt:1987im}. We choose to work with the spatial domain marked by the position $z$, but can equivalently use the temporal domain since we deal with the propagation of relativistic particles in this work. This gives $ i\partial_{z}\textbf{A}=H\textbf{A}$, where the total Hamiltonian $H = H_{0}+H_{1}$ is split into diagonal and off-diagonal components,
\beq\label{Hamiltonian}
   H_{0}= \begin{pmatrix}\w+\Delta&0\\0&\w+\Delta_{\Ap}\end{pmatrix}
   ~,~
   H_{1}=\begin{pmatrix}0&\eps \, \Delta_{\Ap}\\\eps \, \Delta_{\Ap}&0\end{pmatrix}
   ~,
\eeq
with 
\beq
\label{Delta}
\Delta = - \meff^{2}(z) / 2\w
~~\text{and}~~
\Delta_{\Ap}= - \mAp^2 / 2\w
~.
\eeq
Since $\eps \ll 1$, we can approximate $H_1 \ll H_0$ and use the techniques of time-dependent perturbation theory to solve for the evolution of $\textbf{A}$. In particular, we switch to the interaction picture where $i\partial_{z}\textbf{A}_\text{int}=H_\text{int} \, \textbf{A}_\text{int}$, such that $\textbf{A}_\text{int}(z)=\mathcal{U}^{\dagger}(z)\textbf{A}(z)$, $H_\text{int} = \mathcal{U}^{\dagger}H_{1}\mathcal{U}$, and $\mathcal{U}$ is defined to be 
\beq
\mathcal{U}(z)=\exp{\left[-i\int^{z}_{z_{i}} dz' H_{0}(z')\right]}
~,
\eeq
such that $z_i$ marks the point at which we fix our initial condition  $\textbf{A} (z_i) = \textbf{A}_\text{int} (z_i)$. Hence, the system evolves as $\textbf{A}_\text{int} (z) = e^{-i \int_{z_i}^z dz^\prime \, H_\text{int} (z^\prime)} \, \textbf{A}(z_i)$, 
which in the Schr\"odinger picture is equivalent to 
\beq
\label{SchEvo}
\textbf{A} (z) = e^{-i \int_{z_i}^z dz^\prime \, H_0 (z^\prime)} \, e^{-i \int_{z_i}^z dz^\prime \, H_\text{int} (z^\prime)} \, \textbf{A}(z_i)
~.
\eeq
The first factor of \Eq{SchEvo} is determined by the definition of $H_0$ in \Eq{Hamiltonian}. To proceed, we evaluate $H_\text{int}$ using the Baker-Campbell-Hausdorff identity, 
\beq
\label{Hint}
    H_\text{int} = \eps\, \Delta_{\Ap}(z) \, \begin{pmatrix}0& e^{i\Phi(z)}\\ e^{-i\Phi(z)}&0\end{pmatrix}
    ~,
\eeq
where we have defined
 \beq
 \label{phase_def}
    \Phi(z) = \int^{z}_{z_{i}} dz' ~\Delta_\text{osc}(z')
    ~~,~~
    \Delta_\text{osc} = \Delta-\Delta_{\Ap}
    ~.
\eeq
This form of $H_\text{int}$ in \Eq{Hint} can be used in the second factor of \Eq{SchEvo}, after expanding to $\order{\eps}$. Up to an irrelevant overall phase, this yields 
\beq
\textbf{A}(z) \propto 
\begin{pmatrix}1 & -i\eps \, c_+ \\ - i \eps ~ e^{i\Phi(z)} \, c_- & e^{i\Phi(z)}\end{pmatrix} \textbf{A}(z_{i}) + \order{\eps^2}~,
\eeq
where we defined $c_\pm = \int^{z}_{z_{i}}dz' ~ e^{\pm i\Phi(z')} \, \Delta_{\Ap}(z')$. 

The probability of conversion between active and sterile states is then given by the square of the off-diagonal elements in the above expression,
\beq
\label{prob_num}
\prob= \eps^2 \, \bigg| \int^{z}_{z_{i}}dz' ~ \Delta_{\Ap}(z') \, e^{i\Phi(z')} \bigg|^{2} + \order{\eps^3}~.
\eeq
In vacuum, there is no in-medium contribution to the active photon, $\Delta = 0$, and the above integral can be performed analytically, yielding the standard result 
\beq
\label{eq:vacuum}
P_{A_{s}\lrarr A_{a}} = 4\eps^{2}\sin^{2}\left(\Delta_{\Ap}\, (z-z_i)/2\right).
\eeq More generally, for $\eps \ll 1$ and $\Delta \neq 0$, one can numerically integrate \Eq{prob_num} in order to calculate the in-medium conversion probability. In this case, the integrand in \Eq{prob_num} is very oscillatory, making it difficult to achieve a high degree of numerical accuracy. Near resonances, however, it is possible to accurately approximate $\prob$ analytically, as discussed in the next Subsection. 

\subsection{Stationary Phase Approximation and the Landau-Zener Formula}
\label{sec:stationary}
\begin{figure*}
    \includegraphics[width=0.99\textwidth]{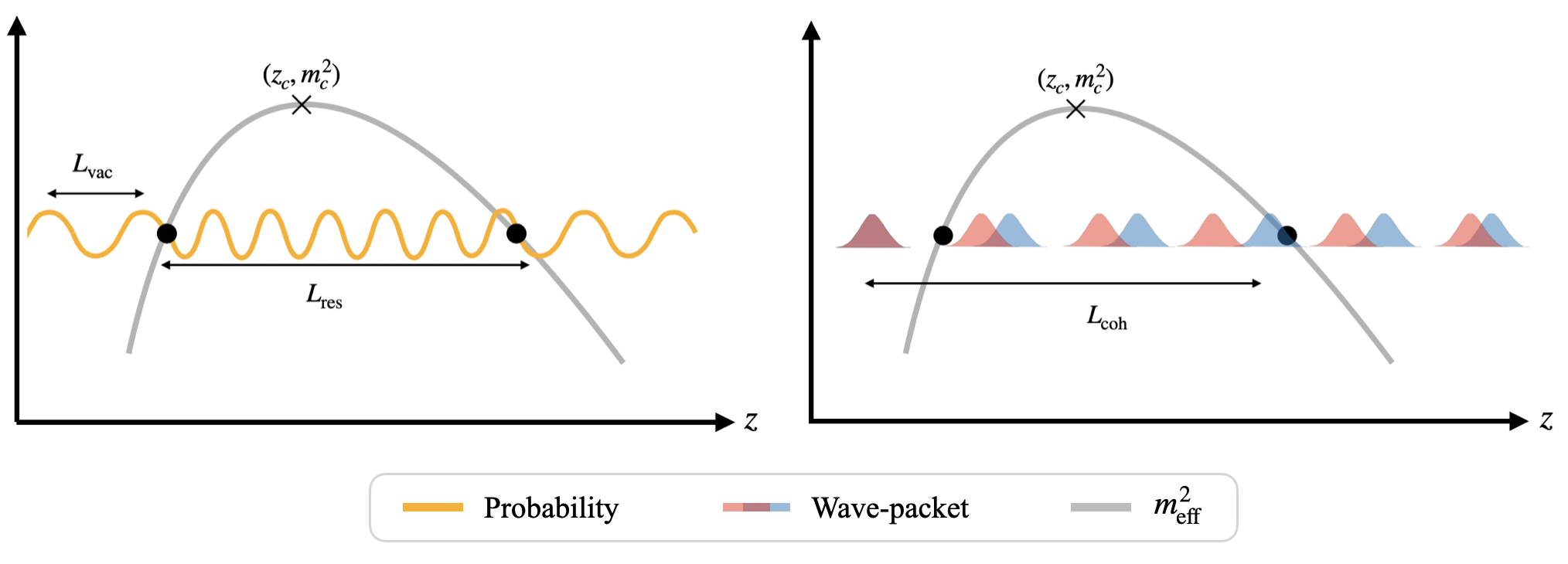}
    \caption{A schematic representation of a peaked $m_{\mathrm{eff}}$-profile (gray line) with the distance $L_{\mathrm{res}}$ between the resonances (black dots) marked for a specific dark photon mass. The critical point is marked by the black ``$\times$" at the top of the potential. Left: A depiction of how the conversion probability oscillates spatially, where $L_{\mathrm{vac}}$ is the relevant conversion length scale in vacuum. Right: Propagating photon and dark photon wave packets are depicted in red and blue, respectively, for the two mass eigenstates, with the coherence length $L_{\mathrm{coh}}$ also shown.}
    \label{potential_plot}
\end{figure*}

While the integrand in \Eq{prob_num} typically exhibits highly oscillatory behavior, it varies slowly near stationary points $z_n$, defined by $\Phi^\prime(z_n)=0$, where the ``prime" corresponds to a spatial derivative. Since the phase varies slowly in this region of coordinate space, its contribution to the integral is not cancelled out by other regions' contributions, where oscillations tend to interfere destructively and average to zero. From \Eq{phase_def}, it is evident that these stationary points occur when the resonance condition holds, $\Delta=\Delta_{\Ap}$, analogous to level-crossings induced by matter effects within the context of neutrino oscillations~\cite{Wolfenstein:1977ue,Mikheyev:1985zog}. In this case, \Eq{prob_num} can be evaluated analytically by use of the stationary phase approximation, in which $\Phi(z^\prime)$ and $\Delta_{\Ap}(z^\prime)$ in the integrand are Taylor expanded to second and zeroth order around $z^\prime \simeq z_n$, respectively. This gives
\begin{align}
\label{stationary_approx}
&\prob \simeq \eps^2 \, \Big| \sum_n \sqrt{A_n} ~ e^{i\Phi(z_{n})+i\sigma_{n}\pi / 4} \Big|^2
\nl
&= \eps^2 \, \Big( \sum_n A_n + 2 \sum_{n<m} \sqrt{A_n \, A_m} \, \cos{\Phi_{nm}} \Big)
~,
\end{align}
where the sums are over all stationary points $z_n \in [z_i, z]$, and in the first and second lines we have defined
\beq
\label{eq:AnDef}
A_n = \frac{2 \pi \, \Delta_{\Ap}^2 (z_n)}{|\Phi''(z_n)|}
~~,~~
\sigma_n = \text{sign} \big[ \Phi''(z_n) \big]
~,
\eeq
and
\beq
\Phi_{nm} = \Phi(z_n) - \Phi(z_m) + (\pi / 4) \, (\sigma_n - \sigma_m)~,
\eeq
respectively, where from \Eq{phase_def} we have $\Phi^{\prime \prime} (z_n) = \Delta^\prime (z_n)$. In the second line of \Eq{stationary_approx}, the first term  is simply the sum of individual probabilities from each resonance. The second term instead stems from the interference between two different resonances, which gives rise to what we will refer to below as \textit{phase effects}. Such phase effects imprint oscillatory behavior into the conversion probability as a function of $\mAp$ and $\w$.

As shown schematically in  \Fig{potential_plot}, the {conversion probability} undergoes a large number of oscillations between any two resonances if $|\Phi_{n m}| \gg 2 \pi$, or equivalently
\beq
\bigg| \int_{z_m(\w)}^{z_n (\w)} dz^\prime ~ \frac{\mAp^2 - \meff^2 (z^\prime,\w)}{2 \w} \bigg| \gg 2 \pi~,
\eeq
where we have been explicit in regards to the spatial and frequency dependence. In this case, small variations in frequency within an observed resolution bandwidth could lead to variations in $\Phi_{nm}$ that are much larger than $2 \pi$, such that the phase in the second term of \Eq{stationary_approx} averages to zero within that frequency bin. A related possibility is if the spatial profile of $\meff(z)$ varies between slightly different lines of sight within the angular resolution of the detector. Again, in the limit where the total acquired phase is large, small variations in the $\meff(z)$ profile can lead to large phase variations within the resolution of the detector. This would potentially cause phase effects to average to zero within an angular bin, depending on the underlying spatial distribution of $\meff$.

An additional factor in the loss of phase information is the coherence length, which determines the ability to maintain phase coherence between different mass eigenstates. This length is the distance over which wavepackets of different mass eigenstates seperate by a distance greater than the wavepacket width, and its dependence on the production process has been investigated extensively in the context of neutrinos~\cite{Nussinov:1976uw,anada1990coherence}. If the distance between resonance regions exceeds the coherence length, then different mass eigenstates decohere before reaching the subsequent resonance region, leading to the loss of phase information. For the purposes of this work, we assume that the states remain coherent, as we are working with relativistic DPs and do not assume that the source of SM photons or DPs is spatially or temporally localized; we will revisit these assumptions and their relevance to astrophysical environments in future work.

In the literature, phase effects are often assumed to average to zero. Ignoring the corresponding cross terms in \Eq{stationary_approx}, the conversion probability reduces to the standard LZ result~\cite{Zener:1932ws,Landau:1932vnv} for a non-adiabatic two-level transition, 
\beq
\label{LZ}
\prob^{\text{LZ}}\simeq \eps^{2} \, \sum_{n}A_{n} ~.
\eeq
Note that this approximation is valid only when the above expression remains less than unity, which would otherwise seemingly violate unitarity. In this work, we always deal with conversion probabilities $\prob \ll 1$ such that multiple sequential conversions are highly suppressed.

\subsection{Breakdown of Landau-Zener}
\label{sec:breakdown}

If $\mAp$ happens to be near a local extremum of $\meff (z)$, there exists a pair of resonant points $z_{1,2}$ such that $z_1 \lesssim z_c \lesssim z_2$ and $z_1 \simeq z_2 \simeq z_c$ (see \Fig{potential_plot}). For a particular value of $\mAp$, such points coalesce near the position of the extremum, $z_{c}$. We refer to this mass $\mAp =\meff(z_{c}) \equiv m_{c}$ as the \textit{critical mass}. At this critical point $z_{c}$, the first two derivatives of $\Phi(z)$ both vanish due to the resonance appearing at the extremum of the $\meff$ potential. Near $z_{c}$, the contributions from the saddle points $A_{n} \propto |\Phi^{\prime \prime}(z_n)|^{-1}$ in \Eq{stationary_approx} diverge. As a result, neither \Eq{stationary_approx} or the LZ approximation in \Eq{LZ} are valid for $\mAp \simeq m_{c}$. 

\begin{figure*}
\includegraphics[width=0.49\textwidth]{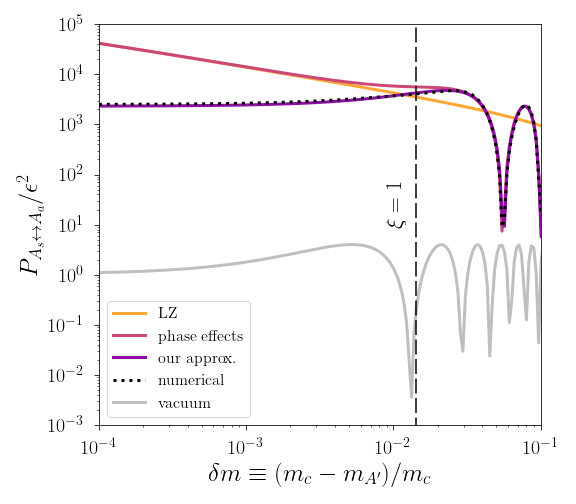}
\includegraphics[width=0.49\textwidth]{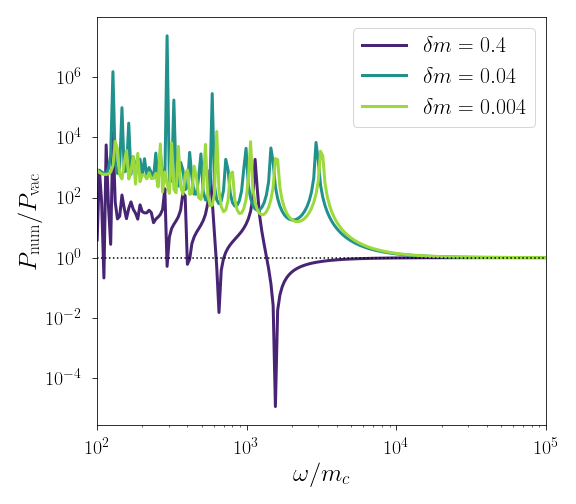}
\vspace{-0.3cm}
        \caption{Left: The conversion probability $\prob$ as a function of $\delta m$ (a dimensionless measure of the proximity  of $\mAp$ to the critical mass $m_c$) for the toy potential of \Eq{toy_potential} with $m_c \, z_c=2\times 10^4$, $\w/m_c = 10^{2}$, and $\w \, z_c = 2\times 10^6$. The region to the left of the dashed vertical line corresponds to $\xi \lesssim 1$ (see \Eq{xidef}), indicating the expected breakdown of standard approximations. The various approximations, consisting of the LZ result of \Eq{LZ}, adding the term that captures phase effects in \Eq{stationary_approx}, and our approximation in \Eq{prob_crit}, are compared to a numerical evaluation of \Eq{prob_num} and the vacuum conversion probability of \Eq{eq:vacuum}.
        Right: The ratio between the full conversion probability computed numerically and the vacuum oscillation probability as a function of $\w/m_c$ for fixed $m_c \, z_c = 2\times 10^4$ and for different values of $\delta m$. For larger frequencies, the conversion probability reduces to its vacuum value as the transition becomes increasingly non-adiabatic.    }
        \vspace{-0.3cm}
        \label{probab_toy}
\end{figure*}

Instead, analytically determining the transition probability when $\mAp \simeq m_c$ requires incorporating the cubic term in the Taylor expansion of $\Phi(z)$ around $z_c$ in \Eq{prob_num}, since $\Phi'(z_c) = \Phi''(z_c) = 0$. We therefore define a dimensionless variable 
\beq
\label{xidef}
\xi \equiv \min_{z_n} \frac{|\Phi^{\prime \prime}(z_n)|}{|\Phi^{\prime \prime \prime} (z_n)|^{\frac{2}{3}}} 
\eeq
that quantifies the relative contribution of the quadratic term over the cubic term in the Taylor expansion. Hence, for $\xi \ll 1$ the quadratic term becomes negligible and one should expand to include $\Phi'''(z_c)$ in order to accurately evaluate the conversion probability near the critical mass. We incorporate this higher-order term by making use of well-known results for two ``coalescing saddle points''~\cite{connor1971theory,beuc2019approximate}, yielding 
\begin{align}
\label{prob_crit}
&\prob \simeq 4\pi^{2} \, \eps^{2} \, \Delta_{\Ap}^{2} \, \bigg(\frac{2}{|\Phi^{\prime \prime \prime}|}\bigg)^{2/3} ~ \bigg[ \text{Ai}\left(-\zeta\right)
\nl
&-i\sigma (z_1) \, \bigg(\frac{2}{|\Phi^{\prime \prime \prime}|}\bigg)^{1/3} \, \bigg(\frac{\w'}{\w}+\frac{1}{6}\frac{\Phi^{\prime \prime \prime \prime}}{\Phi^{\prime \prime \prime
}}\bigg) \, \text{Ai}'(-\zeta)\bigg]^{2}
~ \Bigg|_{z_c}
\end{align}
which is a key result of this work and is only valid for values of $\mAp \simeq m_c$ such that $\xi \lesssim 1$. In \Eq{prob_crit}, the entire expression is evaluated at the critical point $z_c$, $\text{Ai}(x)$ is the Airy function, and $\zeta \equiv \sigma(z_1)\left(2 / |\Phi^{\prime \prime \prime}|\right)^{1/3} \, \Phi^{\prime}$. We note that this formula is valid for two resonant crossings near an extremum; incorporating additional resonance points is possible with the approximations of Ref.~\cite{beuc2019approximate}.

\section{non-monotonic Toy potential}
\label{sec:toymodel}
In this Section, we begin to quantify the degree to which various treatments of DP-SM photon transition probabilities can differ. Motivated by the form of the Taylor expansion of $\meff$ about its extremum,  we consider the following quadratic toy model for the potential,
\beq
\label{toy_potential}
\meff^{2}(z)=m_c^2 \, \bigg[ \, 1-\bigg(\frac{z}{z_c}-1\bigg)^2 ~ \bigg]
~,
\eeq
where we have chosen to take the width of the peak near $z_c$ to be $\mathcal{O}(z_c)$ so that there is only one relevant length scale in the problem. For $\mAp<m_c$, there are two resonant level crossings where $\mAp = \meff$. Because any potential will take a similar quadratic form near its extremum, we can use this toy model as a proxy to gain intuition for potentials in astrophysical systems, such as the ones discussed in the next Section. Conversion probabilities using \Eq{toy_potential} computed with different methods are shown in \Fig{probab_toy} as a function of
\beq
\delta m \equiv (m_c - \mAp) / m_c~.
\eeq

As can be seen in the left panel of \Fig{probab_toy}, when $\mAp$ is far from the critical mass $m_{c}$, the approximation in \Eq{stationary_approx} (labelled ``phase effects") matches well with the numerical evaluation of \Eq{prob_num} (labelled ``numerical"). In this case, the standard LZ result of \Eq{LZ} (labelled ``LZ") accurately captures the typical value of the transition probability, averaged over the oscillatory features with varying $\mAp$. As $\mAp$ approaches $\meff$ from below, the two resonance points converge spatially, eventually merging at the critical point when $\delta m = 0$, corresponding to $\xi \to 0$ in \Eq{xidef}. As discussed in the previous Section, the approximations detailed in \Eqs{stationary_approx}{LZ} are no longer valid for $\xi \lesssim 1$ (left of the vertical dashed line). Nevertheless, our approximation in \Eq{prob_crit} (labelled ``our approx.") remains accurate and proves to be a reliable method of tracking the conversion probability near this critical mass. We emphasize that these approximations are substantially faster to evaluate numerically as compared with the full numerical solution of the Schr\"odinger equation, primarily due to the oscillatory nature of $\Phi(z)$.

In-medium resonances significantly amplify the conversion probability with respect to the vacuum value of \Eq{eq:vacuum}. To encapsulate the enhancement due to such in-medium resonances, we first note that \Eq{eq:AnDef} can be rewritten as $A_n = 2 \pi \, \mu_n $, where we have introduced a dimensionless \mutitle
\beq
\label{eq:mun}
\mu_n = \Big| \Delta_{\Ap} (z_n) \Big| \times \Big|\frac{d}{dz} \log{|\Delta(z_n)|} \Big|^{-1}
\eeq
that is analogous to the adiabaticity parameter of Keldysh in Ref.~\cite{Keldysh:1965ojf}. Since $\prob \sim \eps^2 \max{A_n}$ in the LZ approximation, $\mu \equiv \max \mu_n$ effectively quantifies the degree that any such resonance enhances the conversion probability over the vacuum value. Hence, for $\mu \ll 1$, we expect the in-medium modifications to the transition probability to be suppressed, such that $\prob$ approaches the vacuum value of \Eq{eq:vacuum}. 

This form of the \mutitle $\mu$ is to be expected. To see this, note that for a non-monotonic potential, such as the one of \Eq{toy_potential}, $\mu_n \sim L_\text{res} / L_\text{vac}$, where $L_\text{res}$ is the separation between $z_n$ and a nearby resonant point, and $L_\text{vac} \sim 1 / \Delta_{\Ap}$ is the vacuum oscillation length. Hence, for $\mu \ll 1$, the potential as seen by the DP undergoes a spatially abrupt change compared to the oscillation length, resulting in an extremely non-adiabatic transition where it is valid to use the ``sudden approximation.'' Since by the uncertainty relation the DP can only resolve length scales greater than $1/\Delta_{\Ap} \sim L_\text{vac}$ for a momentum difference between active and sterile states of $\delta k = \Delta_{\Ap}$, it cannot effectively discern the presence of the potential over a distance of $L_\text{res} \ll L_\text{vac}$. Conversely, for $\mu \gg 1$, the potential varies over long distances compared to the vacuum oscillation length, indicating a need to account for $\meff$ using one of the other approximations detailed in the previous Section.

The validity of the sudden approximation (i.e., using the vacuum transition probability) for $\mu \ll 1$ is verified numerically in the right panel of \Fig{probab_toy}, which compares numerical evaluations of $\prob$ with \Eq{prob_num} to the vacuum value in \Eq{eq:vacuum}, as a function of $\omega$ (expressed in terms of the dimensionless quantity $\omega/m_c$) and for various representative choices of the DP mass in the form of $\delta m$. As $\w$ increases, the {vacuum} oscillation length also increases compared to the effective width of the $\meff$ profile, such that the conversion probability progressively approaches its vacuum value. 

\section{astrophysical and cosmological environments}
\label{sec:astro}
In this Section, we provide a few relevant examples of environments with non-monotonic effective photon masses where it is necessary to use Eq.~\eqref{prob_crit} in order to accurately evaluate the DP conversion probability near the critical mass. It may be necessary to update some astrophysical and cosmological bounds on DPs for particular DP masses in light of these considerations. 

\subsection{Neutron Star Magnetospheres}
\begin{figure*}[t]
\includegraphics[width=0.49\textwidth]{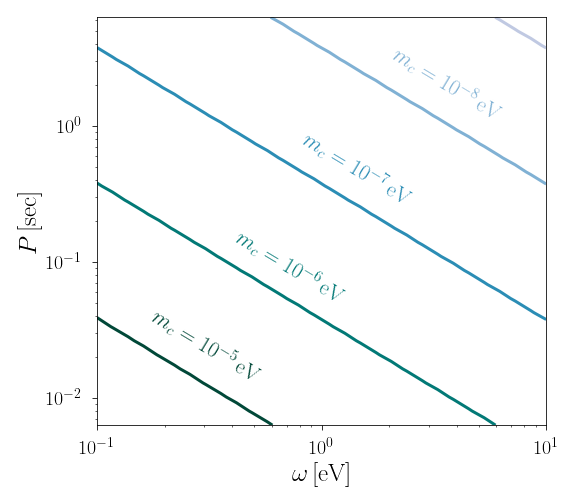}
    \includegraphics[width=0.49\textwidth]{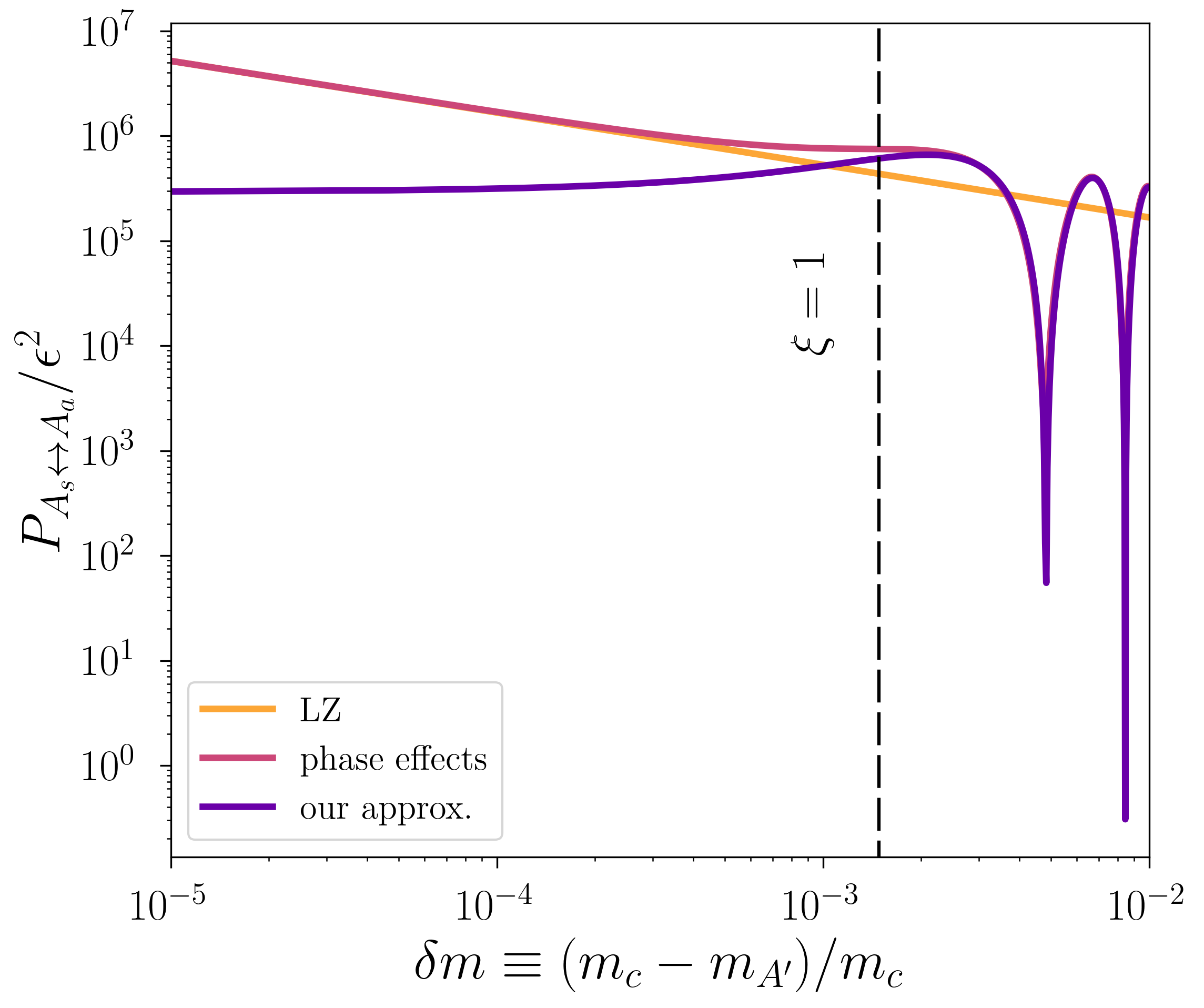}
    \vspace{-0.3cm}
    \caption{Left: Contours of the critical mass $m_c$ in the environment of a neutron star, in the plane spanned by  the rotation period $P$ of the NS and frequency $\w$ for a fixed value of the magnetic field $B_0/B_c = 10$. The critical mass varies considerably for fixed neutron star properties, meaning that any conversion processes occurring over a range of frequencies would only be accurately described by the approximations developed in this work. Right: As in \Fig{probab_toy}, but for the neutron star potential of \Eq{NS_potential} with $r_{\mathrm{LC}} = 300~ \mathrm{km}$, $B_0/B_c=10$, and $\w = 0.08 \ \eV$. The similarity of this behavior with that of the toy model underscores the necessity of using the approximation near the critical mass.}
    \vspace{-0.3cm}
    \label{NS_plots}
\end{figure*}

Among astrophysical environments, neutron star (NS) magnetospheres are the most extreme in their variations in free charge density and electromagnetic field strength; they therefore pose an environment where $\meff^2$ can vary substantially, affecting conversion of DPs to SM photons~\cite{Fortin:2019npr}. In particular, the rotating magnetic fields source electric fields that exert forces much larger than the gravitational binding energy on the NS surface charges. As a result, charges are pulled off of the surface and fill the magnetosphere of the NS.  The charges then redistribute themselves in a way such that the Lorentz force on them cancels, producing a corotating plasma surrounding the NS. The charge density of this plasma can be approximated by the Goldreich-Julian model~\cite{goldreich1969pulsar}, which has been generalized to account for relativistic effects~\cite{Mofiz:2000xsn}
\begin{align}
n_{e}^{\text{GJ}}(r,\theta) &= \frac{2 \, \mathbf{\Omega
}\cdot\textbf{B}}{e} \, \Big[F_{1}(\bar{r}) \, \sin^{2}\theta 
\nl
& -F_{2}(\bar{r}) \, \left(\sin^{2}\theta-2\cos^{2}\theta\right)\Big]
~. \label{eq:GJ}
\end{align}
Here, $r$ and $\theta$ are polar coordinates such that $r$ is the distance from the center of the NS, $\theta$ is the polar angle with respect to the NS's rotation axis $\hat{\mathbf{\Omega}}$, $\textbf{B}$ is the magnetic field outside the NS (assumed to be dominated by its dipole component), and  $\Omega$ is the rotational frequency. The functions 
\begin{align}
& F_{1}(\bar{r})=\bar{r}^{3} \, \Bigg\{ \bigg(1-\frac{\beta}{\bar{r}^{3}}\bigg)\bigg[ \frac{2}{\bar{r}-1}-\frac{1}{(\bar{r}-1)^{2}}
+2\ln\Big(1-\frac{1}{\bar{r}}\Big)\bigg] \Bigg\}
\nl 
&+ \bigg(2+\frac{\beta}{\bar{r}^{3}}\bigg)\Bigg\{ \frac{1}{\bar{r}}+\frac{1}{\bar{r}-1}+2\ln\Big(1-\frac{1}{\bar{r}}\Big)\Bigg\} 
\end{align}
and
\begin{align}
F_{2}(\bar{r})=\bar{r}^{3}\frac{2\left(1-\frac{\beta}{\bar{r}^{3}}\right)}{1-\frac{1}{\bar{r}}}\Bigg\{ \frac{1}{2\bar{r}^{2}}+\frac{1}{\bar{r}}+\ln\Big(1-\frac{1}{\bar{r}}\Big)\Bigg\} 
\end{align}
incorporate relativistic corrections, where we have defined $\bar{r}=r/r_{g}$ and $\beta=2/5(r_{\mathrm{NS}}/r_{g})^{2}$ for a NS of radius $r_{\mathrm{NS}}$ and Schwarschild radius of $r_{g}$. Note that at large distances, $\lim_{\bar{r} \to \infty} F_1(\bar{r}) = -1/2 \bar{r}$ and $\lim_{\bar{r} \to \infty} F_2(\bar{r}) = -2/3$.

Following Ref.~\cite{Hook:2018iia}, we take the rotation axis to be aligned with $\hat{\textbf{z}}$, such that
\beq
\mathbf{\Omega} \cdot \mathbf{B}=\frac{B_{0}}{2r_{\mathrm{LC}}} \, \left(\frac{r_\text{NS}}{r}\right)^{3} \, \left[3 \, \cos\theta\, \mathbf{\hat{m}} \cdot\mathbf{\hat{r}}-\cos\alpha_{B}\right]
\eeq
and
\beq
\mathbf{\hat{m}} \cdot \mathbf{\hat{r}}=\cos\theta\cos\alpha_{B}+\sin\theta\sin\alpha_B \cos \left(\Omega t\right)
~,
\eeq 
where $B_{0}$ is the magnetic field strength at the NS surface, $\alpha_B$ is the orientation angle of the magnetic field with respect to the rotation axis, and $r_\text{LC}=1/\Omega$ is the light-cylinder radius.  

In the presence of large magnetic fields (with $B_0\sim B_c$ where $B_c=m_{e}^{2}/e$ is the critical value of the magnetic field), both the magnetic field and the plasma contribute to $\meff$ for the propagating photon mode (that has its electric field parallel to the plane containing the propagation and NS magnetic field vectors) but with opposite signs~\cite{Lai:2001di, Potekhin:2004jr, Fortin:2019npr}. We can decompose the effective SM photon mass as $m_{\mathrm{eff}}^2=V_{B}+V_{\mathrm{pl}}$, where
\beq
\label{NS_potential}
V_{B}\simeq -\frac{7\alpha}{45\pi} \, b^{2} \, \hat{q}_B \, \w^{2} \, \sin^{2}\theta
~~,~~
V_{\mathrm{pl}}= \w_{p}^{2} \, \sin^{2}\theta
~,
\eeq
$b=B(r)/B_c$,  $\w_{\mathrm{pl}}^{2} = 4\pi\alpha n_{e}^{\text{GJ}}/m_e$ is the plasma frequency, and $\hat{q}_B$ is a fitting function that reproduces the correct $b\ll 1$ and $b\gg 1$ behavior~\cite{Potekhin:2004jr}
\beq
\hat{q}_B \equiv \frac{1+1.2b}{1+1.33b+0.56b^2}
~.
\eeq
The frequency-dependent $B$-field contribution to the photon potential $V_B$ originates from non-linear vacuum polarization effects (analogous to the Euler-Heisenberg term in pure QED) and contributes effectively only when $B(r)\gtrsim B_c$. The opposing signs of $V_B$ and $V_\text{pl}$ give rise to a non-monotonic profile for $\meff^2 (r)\, $. Note that the potential is non-monotonic regardless of the inclusion of relativistic corrections in \Eq{eq:GJ}.

The exact form of the SM photon potential $m_\text{eff}^2 (r)$ is sensitive to various NS parameters, such as the magnetic field strength, rotation speed, and frequency $\w$, resulting in a wide range of possible profiles. Since the electron density and magnetic field both scale as $\propto 1/r^3$ far from the NS's surface, the potentials of \Eq{NS_potential} scale as $V_B \propto - r^{-6}$ and $V_\text{pl} \propto r^{-3}$. As a result, $\meff^2 < 0$ when $V_B$ dominates at intermediate distances (which precludes the possibility of a resonance in this region) and 
$\meff^2 > 0$ when $V_\text{pl}$ dominates at large distances. Hence, the potential reaches its extremum at the turnover point $r_c$ when $V_\text{pl}(r_{c})\sim -V_{B}(r_{c})$. For $B_0 \lesssim B_c$, the corresponding critical radius and mass take simple analytical forms,
\beq
r_{c}  \sim \big( \w^2 \, r_\text{LC} \, B_0 \big)^{1/3} \, \frac{e \, r_\text{NS}}{m_e}
~ , ~
m_{c}  \sim 
\frac{m_e}{e \, \w \, r_\text{LC}}\, .
\eeq

As discussed in \Sec{sec:breakdown}, when the dark photon is close to but slightly less than the critical mass, there exist two nearby resonant points $r_1 \simeq r_2$ such that $r_1 \lesssim r_c \lesssim r_2$. These resonances are physically realized only if the critical radius $r_c$ lies in the range $r_{\mathrm{NS}} \leq r_{c} \leq r_\text{LC}$, which is equivalent to 
\beq
\label{NS_bounds}
\frac{r_{\mathrm{NS}}^{2}}{r_\text{LC}} 
\lesssim
\frac{e^2}{m_e} \, \frac{B_0}{B_c} \, \w^{2} \lesssim \frac{r_\text{LC}^{2}}{r_\text{NS}} 
~.
\eeq
The left panel of \Fig{NS_plots} shows that the critical mass $m_c$ spans many orders of magnitude for a representative range of possible values for the light-cylinder radius $r_\text{LC}$ and frequency $\w$, fixing $B_0 / B_c = 10$ and using the full form for the potential $\meff^2$. In the right panel, we show the conversion probability as a function of $\delta m$, using the various approximations discussed in \Sec{sec:oscill}, for a particular choice of NS parameters. By comparing to \Fig{probab_toy}, we note that our findings qualitatively resemble the probabilities for the toy example of the previous Section, further illustrating that \Eq{prob_crit} is required to accurately compute $\prob$ for $\delta m \ll 1$.

\subsection{Intergalactic Medium}

\begin{figure*}
    \includegraphics[width=0.49\textwidth]{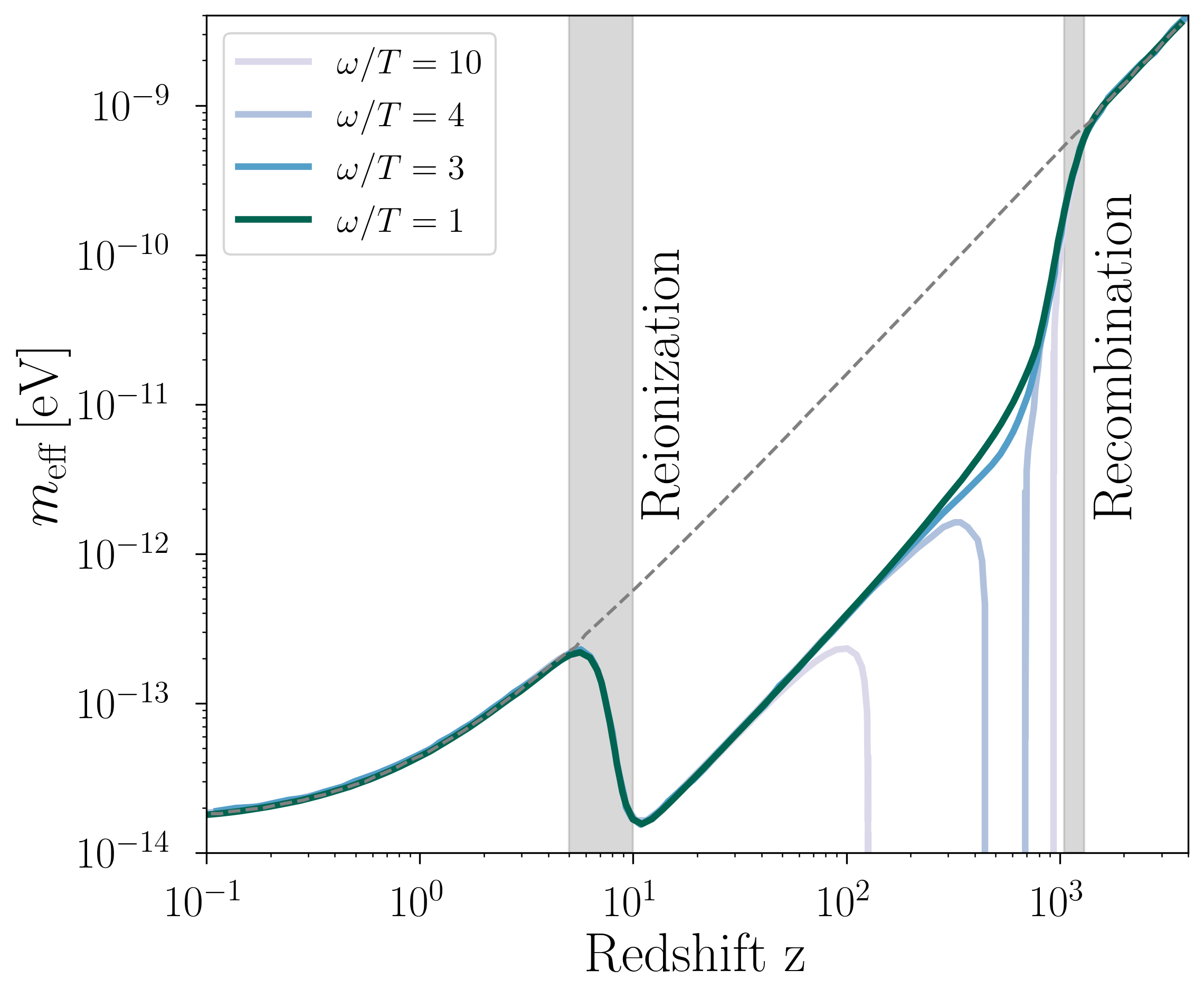}
    \includegraphics[width=0.475\textwidth]{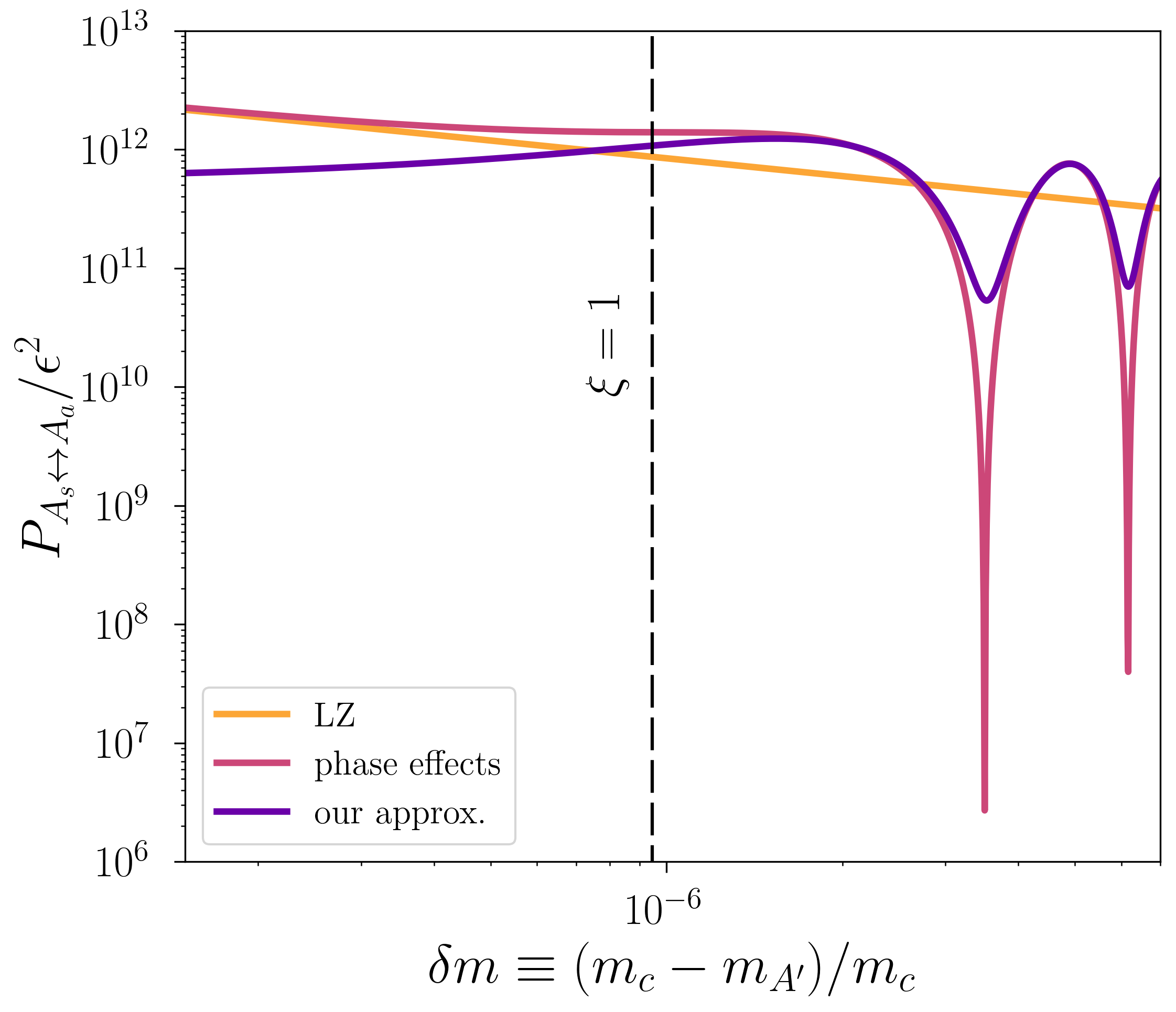}
    \\
    \includegraphics[width=0.49\textwidth]{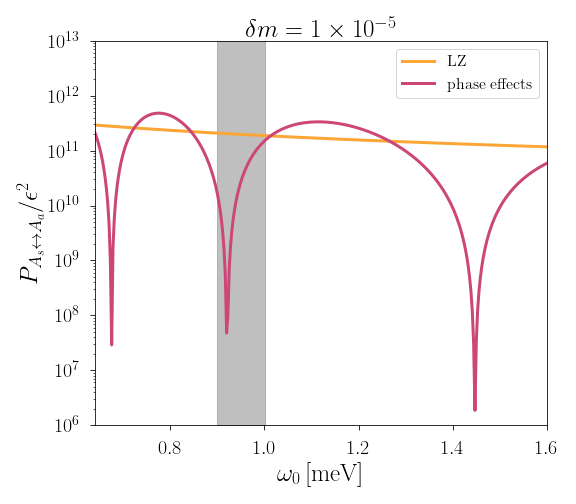}
    \includegraphics[width=0.49\textwidth]{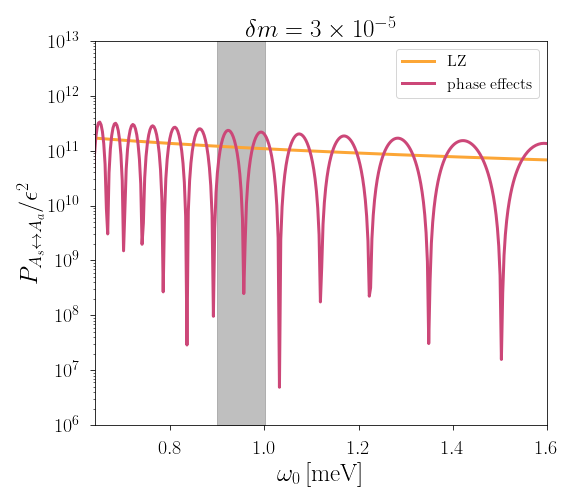}
    \vspace{-0.3cm}
    \caption{Top left: The spatially averaged evolution of $\meff$ in the intergalactic medium as a function of redshift for different frequencies. Reionization induces extrema in the potential experienced by CMB photons. Top right: As in \Fig{probab_toy}, but for the IGM potential shown in the top-left panel for a frequency of $\omega = 7\times 10^{-4}$ eV and $m_c \simeq 2.5 \times 10^{-13}$ eV. Bottom: As in \Fig{probab_toy}, but with two values of fixed dark photon masses and instead varying the frequency. Also shown as a grey vertical band is the frequency resolution of FIRAS, 24.6 GHz \cite{firasnasa}. For values of the dark photon mass near the critical mass, the frequency oscillations do not necessarily average out within the frequency resolution of the instrument, meaning that phase effects are potentially significant in determining the observed conversion probability. Note that in the bottom panels, the values of $\delta m$ correspond to $\xi>1$ where we expect our approximation is not necessary to compute the conversion probability.}
    \vspace{-0.3cm}
    \label{cosmo_plot}
\end{figure*}

The free electron fraction of the intergalactic medium (IGM) has fluctuated considerably since recombination, diminishing during the dark ages and cosmic dawn before increasing again during reionization, all while the density of the Universe was redshifting as $\sim (1+z)^3$. As a result, cosmic microwave background (CMB) photons experience large variations in their effective mass as they traverse the IGM, 
\beq
\meff^2 \simeq \w_{p,e}^2 - (\w / \text{Ry})^2 \, \w_{p, \text{HI}}^2
~,
\eeq
where $\w_{p,e} = \sqrt{4 \pi \alpha \, n_e / m_e}$ and $\w_{p,\text{HI}} = \sqrt{4 \pi \alpha \, n_\text{HI} / m_e}$ are contributions to the plasma frequency from the average number density of free electrons $n_e$ and electrons bound in neutral hydrogen $n_\text{HI}$, respectively. As shown in the top-left panel of \Fig{cosmo_plot}, the resulting SM photon potential has two local extrema independent of frequency near $z \sim 10$, corresponding to just before and after reionization, as well as additional local maxima for frequencies significantly larger than the CMB blackbody temperature, i.e., $\w / T \gtrsim \text{few}$ ~\cite{Mirizzi:2009iz,Seager:1999bc}
. The corresponding values of the critical mass are roughly $m_c \sim \text{few} \times 10^{-13} \ \eV$, $m_c \sim 10^{-14} \ \eV$, and $m_c \sim 10^{-11} \ \eV \times (T/\w)^2$. We note that the first two of these are subject to uncertainties in the ionization history. We postpone a more careful consideration of alternative parametrizations of reionization~\cite{Greig:2016wjs} as well as the incorporation of fluctuations in the plasma density (along the lines of Refs.~\cite{Caputo:2020bdy, Caputo:2020rnx}) to future study. 

\begin{figure*}[t!]
    \includegraphics[width=0.49\textwidth]{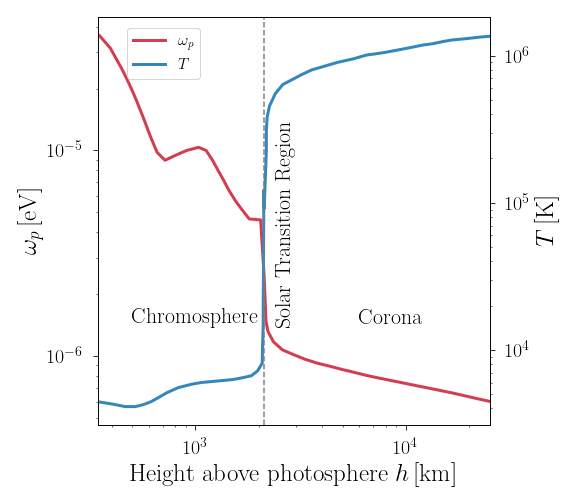}
    \includegraphics[width=0.49\textwidth]{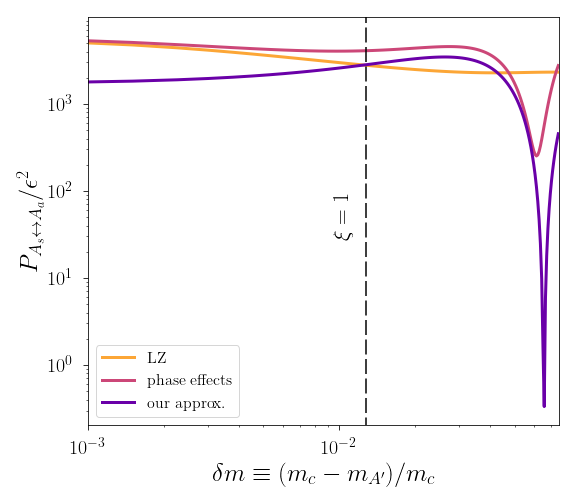}
    \vspace{-0.3cm}
    \caption{Left: The non-monotonic plasma frequency $\w_p$ and temperature $T$ profile above the solar photosphere. The solar transition region which separates the chromosphere and corona is also shown. Right: As in \Fig{probab_toy}, but with the chromospheric potential depicted in the left panel.}
    \vspace{-0.4cm}
    \label{solar_plot}
    \end{figure*}

Previous studies have derived strong constraints on DPs from requiring that the CMB spectrum remains a blackbody, as resonant conversion from CMB photons to DPs could induce obervable spectral distortions in the frequency range probed by the Far InfraRed Absolute Spectrophotometer (FIRAS)~\cite{Fixsen:1996nj}. Accounting for both the mean value and fluctuations of the plasma density has led to strong constraints at the level of $\eps \lesssim 10^{-7}$ for $10^{-15} \ \eV \lesssim \mAp\lesssim 10^{-6} \ \eV$~\cite{Mirizzi:2009iz,Caputo:2020bdy, Caputo:2020rnx,Garcia:2020qrp}. Such studies have employed the LZ approximation discussed above, but this is expected to break down for $\mAp \simeq m_c$, necessitating the use of \Eq{prob_crit}. This is demonstrated in the top-right panel of \Fig{cosmo_plot}, which shows the conversion probability for DP masses near the critical point as computed using different approximation schemes, analgous to the toy example of \Fig{probab_toy}. Given the form of the mean photon potential (i.e., ignoring plasma density fluctuations), DPs with $10^{-13} \ \eV \lesssim \mAp \lesssim 10^{-14}\ \eV$ have three or more resonance points, with two critical points occurring around the time of reionization. It is noteworthy that both $\w(t)$ and $\Phi^{\prime \prime}(t)$ decrease steeply with the age of the Universe, meaning that resonances that occur later are more adiabtic. As a result {of the $\sim1/(\omega^2 \Phi^{\prime \prime})$ scaling of the $A_n$ in \Eq{eq:AnDef}}, resonances occurring at later times therefore have a much greater contribution to the total conversion probability as computed with the LZ formalism of \Eq{LZ}. We therefore ignore the earliest level crossings that occur well before reionization in computing the conversion probability and include only the ones occurring around the time of reionization. 

The conversion probability fluctuates sharply as a function $\w$ and $\meff$ due to the phase effects discussed in \Sec{sec:stationary}. Effects from variations in $\w$ are potentially observable depending on the frequency resolution of the detector, which is shown for FIRAS as the vertical gray region in the bottom row of \Fig{cosmo_plot}. Variations in $\meff$ arise from the fact that different lines of sight trace slightly different ionization histories due to the patchy morphology of reionization, leading to shifts in the SM photon potential and conversion probability along slightly different lines of sight. We note that this opens the possibility for the constraints on DPs from CMB spectral distortions to form more of a ``fog'' (due to theoretical uncertainties in computing the conversion probabilities) rather than a sharp exclusion boundary in parameter space. We leave the question of whether or not these variations average out across different lines of sight to future work. We also note that these considerations may be relevant for the dark screening effect that was recently proposed in Ref.~\cite{Pirvu:2023lch}, since there the photons passing through halos containing a gas overdensity will have two resonance points near the maximum plasma frequency. 

\subsection{Solar chromosphere}
In stellar chromospheres, the density of free electrons $n_e^c$ exhibits a distinct peak located approximately $10^3 \ \km$ above the photosphere, as depicted in \Fig{solar_plot}~\cite{aschwanden2006physics}. This non-monotonic electron density translates to a non-monotonic $\meff^2 =4 \pi \alpha (n_e^c)^2/m_e$. Additionally, the form of the profile yields a small value of $\xi$ for a wide range of frequencies if the DP mass is near the critical mass. For instance, for $\delta m \sim 0.01$, $\xi \lesssim 1$ for $\w \gtrsim0.1 \ \eV$. This implies the unavoidable need for the approximation of \Eq{prob_crit} to accurately assess the conversion probability shown in \Fig{solar_plot}. This may be important to incorporate for DP searches involving the Sun, for instance strategies that could complement existing searches involving resonant level crossing in the solar atmosphere (e.g., Ref.~\cite{An:2023wij}). We explore this possibility in future work. 

\section{Conclusion}
\label{sec:conclusion}
In this paper, we have developed an accurate approximation for the solution to the Schr\"odinger equation for a two-state system (specifically, photons and dark photons) with multiple nearby resonances about the extremum of a potential. This approximation can be viewed as an extension of the LZ formula, which is widely used for computing the transition probability between photons and dark photons. Using a toy model, we find that there can be large corrections to the conversion probability for specific dark photon masses in a given potential (i.e., given some properties of the background environment that affect the propagation of photons). 

We have highlighted the application of this formalism to various astrophysical systems where multiple resonances between dark photons and photons are possible. Some of these systems have been previously used to constrain the existence of dark photons through their observed spectral signatures, which makes it important to quantify the effects of multiple resonances on the constraints. We have not attempted to update any of these constraints from specific astrophysical systems, and leave such an analysis to future work. We note that applying our formalism to neutron stars seems especially promising due to the large resonant enhancement factors over the vacuum conversion probability as well as the wide range of possible values of the critical mass.

There are several additional subtleties in the conversion between photons and dark photons that we have not considered here, such as the role of decoherence between different states, which may be particularly relevant when the dark photon is non-relativistic (e.g., in the case of dark photon dark matter). We note that much of the formalism here, along with the associated subtleties, may be transferred over to axion-photon conversions and neutrino oscillations. We leave consideration of all of these effects to future work. 

\section*{Acknowledgements}
It is a pleasure to thank Bryce Cyr, Adrian Liu, Hongwan Liu, and Anirudh Prabhu for valuable conversations and correspondence pertaining to this work. The research of NB was undertaken thanks in part to funding from the Canada First Research Excellence Fund through the Arthur B. McDonald Canadian Astroparticle Physics Research Institute. NB and KS acknowledge support from a Natural Sciences and Engineering Research Council of Canada Subatomic Physics Discovery Grant and from the Canada Research Chairs program. AB is supported by the U.S. Department of Energy, Office of Science, National Quantum Information Science Research Centers, Superconducting Quantum Materials and Systems Center (SQMS) under contract number DE-AC02-07CH11359. Fermilab is operated by the Fermi Research Alliance, LLC under Contract DE-AC02-07CH11359 with the U.S. Department of Energy. Research at Perimeter Institute is supported in part by the Government of Canada through the Department of Innovation, Science and Economic Development Canada and by the Province of Ontario through the Ministry of Colleges and Universities.
\bibliography{main}
\end{document}